\documentclass[12pt]{article}

\usepackage[utf8]{inputenc}
\usepackage[T1]{fontenc}
\usepackage{times}
\usepackage{geometry}
\usepackage{amsmath, amssymb}
\usepackage{graphicx}
\usepackage{hyperref}
\usepackage{xcolor}
\usepackage{booktabs}
\usepackage{authblk}
\usepackage{setspace}
\usepackage[numbers,sort&compress]{natbib}
\usepackage{lineno}
\usepackage[normalem]{ulem}
\usepackage{amssymb}
\usepackage{pifont}
\usepackage{subcaption}  %\borja{[for sub-panels (a)(b)(c) in the QPU realisation figure]}

\hypersetup{
  colorlinks=true,
  linkcolor=blue,
  citecolor=blue,
  urlcolor=blue
}
\begin{document}

%% ---- TITLE ----
\title{\textbf{Quantum-enhanced Large Language Models on\\
Quantum Hardware via Cayley Unitary Adapters}}

%% ---- AUTHORS ----
\author[1,2]{Borja Aizpurua}
\author[3]{Sukhbinder Singh}
\author[1]{Augustine Kshetrimayum}
\author[1,4]{Saeed S. Jahromi}
\author[1,4,5,*]{Román Orús}

\affil[1]{Multiverse Computing, Parque Científico y Tecnológico de Gipuzkoa,
  Paseo de Miramón 170, 20014, Donostia / San Sebastián, Spain}
\affil[2]{Department of Basic Sciences, Tecnun -- University of Navarra,
  San Sebastián, Spain}
\affil[3]{Multiverse Computing, Centre for Social Innovation,
  192 Spadina Avenue, ON M5T 2C2, Toronto, Canada}
\affil[4]{Donostia International Physics Center,
  Paseo Manuel de Lardizabal 4, 20018, San Sebastián, Spain}
\affil[5]{Ikerbasque Foundation for Science,
  Maria Diaz de Haro 3, 48013, Bilbao, Spain}
\affil[*]{Correspondence: \texttt{roman.orus@multiversecomputing.com}}

\date{}
\maketitle

\noindent\rule{\textwidth}{0.4pt}
\vspace{0.3em}

%% ================================================================
%% SUMMARY PARAGRAPH (bold, ≤150 words, no references, no equations)
%% ================================================================

\noindent\textbf{%
Large language models (LLMs) have transformed artificial intelligence, yet
classical architectures impose a fundamental constraint: every trainable
parameter demands classical memory that scales unfavourably with model size.
Quantum computing offers a qualitatively different pathway, but practical
demonstrations on real hardware have remained elusive for models of
practical relevance.
Here we show that Cayley-parameterised unitary adapters---quantum circuit
blocks inserted into the frozen projection layers of pre-trained LLMs and
executed on a 156-qubit IBM Quantum System Two superconducting
processor---improve the perplexity of Llama~3.1~8B, an
8-billion-parameter model in widespread use, by 1.4\% with only 6,000
additional parameters and end-to-end inference validated on real \emph{Quantum Processing Unit (QPU)}.
A systematic study on SmolLM2 (135M parameters), chosen for its tractability,
reveals monotonically improving perplexity with unitary block dimension,
83\% recovery of compression-induced degradation, and correct answers to
questions that both classical baselines fail---with a sharp noise--expressivity
phase transition identifying the concrete path to quantum utility
at larger qubit scales.%
}

\noindent\rule{\textwidth}{0.4pt}
\vspace{1em}

%% ================================================================
%% MAIN TEXT — begins without an "Introduction" heading (Nature style)
%% ================================================================

We stand at a historical juncture in artificial intelligence.
Large language models (LLMs) have redefined the boundary of machine
intelligence, enabling systems capable of complex reasoning, code synthesis,
and scientific inference~\cite{brown2020gpt3,llama3_2024}.
Yet this progress rests on a substrate approaching a fundamental physical
limit: every trainable parameter occupies classical memory, and scaling
a deployed model's parameter count requires a proportionally unsustainable
expansion of compute infrastructure~\cite{kaplan2020scaling}.
Proposed mitigations---quantisation, pruning, and low-rank adaptation
(LoRA)~\cite{hu2022lora}---reduce footprint at the cost of expressive
capacity.
Tensor-network (TN) methods~\cite{novikov2015tensorizing,orus2014practical,
orus2019tensor} offer structured compression, but their parameter count
remains bounded by classical memory, and the bond dimension $\chi$ must be
stored explicitly.

Quantum computing operates under a fundamentally different resource paradigm. An $n$-qubit system resides in a Hilbert space of dimension $2^n$, while the required physical resources scale only linearly with $n$. Quantum circuits provide a structured family of unitary transformations on this exponentially large space, with several avenues for classical optimisation. Variational quantum circuits (VQCs)—unitaries parameterised by gate-level rotation angles and trained on hardware, for example via the parameter-shift rule~\cite{cerezo2021variational}—offer a widely used mechanism for incorporating quantum parameters into (quantum) machine learning models, including within transformer architectures. In practice, VQCs typically adopt fixed ansätze composed of repeated layers of parameterised single-qubit rotations interleaved with non-variational two-qubit entangling gates, such as CNOTs.

Prior work has explored quantum machine learning approaches to large language models (LLMs) in restricted regimes, including classification~\cite{havlicek2019supervised}, quantum natural language processing on toy grammatical structures~\cite{coecke2020foundations}, and variational sequence models~\cite{bausch2020recurrent}. More recently, hybrid methods for LLM fine-tuning have been proposed~\cite{yu2025qllm,li2025quantum}, and quantum self-attention has been demonstrated on a 72-qubit processor for text classification~\cite{zhao2024selfattention}. Additional advances include quantum knowledge distillation~\cite{quantum_kd2025}, multi-architecture frameworks for quantum-enhanced natural language generation~\cite{qenhanced_nlg2025}, and reinterpretations of transformer layers as unitary operators~\cite{qlens2025}. However, these approaches are limited to simulators, focus on classification rather than autoregressive generation, or operate at restricted linguistic scale. To the best of our knowledge, no prior work has demonstrated quantum enhancement of a production-scale, pre-trained LLM for autoregressive language generation on real gate-based quantum hardware.

Here we address this gap by introducing a simple and hardware-efficient strategy for injecting quantum parameters into classical LLMs. Rather than employing standard VQCs, we adopt a complementary construction based on block-diagonal unitaries (BDU). Each unitary block is parameterised via the Cayley transform of a skew-symmetric matrix, with the upper-triangular components of these matrices, together with the associated Cayley parameters, forming the set of variational degrees of freedom. The number of blocks is determined by the residual stream dimension $d$ of the LLM once the block size is fixed. After insertion at an appropriate location within the model, these parameters are trained entirely classically—avoiding the overhead of gradient evaluation on large datasets—while all original model weights remain frozen. The resulting hybrid model is subsequently executed on quantum hardware.

A key feature of this ansatz is its hardware efficiency. A generic $d \times d$ unitary acting on $n = \log_2 d$ qubits would require an exact synthesis whose depth grows exponentially with $n$, far beyond present-day coherence budgets. Our BDU construction sidesteps this: although the trained adapter is structurally entangling at the global level (rank-saturating across its natural block-index $\mid$ intra-block bipartition; Extended Data Table~9), it factorises into independent $2^n \times 2^n$ blocks executed in parallel as shallow $n$-qubit circuits. For all QPU experiments in this work we fix the block size to $4 \times 4$, so each block is a 2-qubit unitary --- a depth-19 native-gate circuit on \texttt{ibm\_basquecountry}, well within coherence limits. The systematic SmolLM2 study additionally sweeps the block size up to the full input dimension to characterise the expressivity ceiling. This construction provides a practical and scalable route to quantum enhancement of contemporary LLMs.

Using this approach we present two complementary results.
Our \emph{primary result} is the quantum enhancement of Llama~3.1~8B, one
of the most widely deployed open-weight LLMs in current use with
8~billion parameters.
By inserting Cayley-parameterised block-diagonal unitary adapters---quantum
circuit blocks executed on a 156-qubit IBM Quantum System
Two~\cite{ibm_system_two}---we improve WikiText perplexity by 1.4\% with
only 6,000 additional parameters, with end-to-end inference validated
on real Quantum Processing Unit (QPU).
We consider this a foundational result, analogous to the experimental realisation of Shor's algorithm by Vandersypen \textit{et al.} via NMR~\cite{vandersypen2001}, in that it demonstrates the viability of the underlying physical approach and establishes a concrete basis for future scaling.

Our \emph{complementary result} is a systematic mechanistic study using
SmolLM2~\cite{smollm2024} (135M~parameters), a compact but well-characterised
model that is substantially more tractable for exhaustive experimentation.
SmolLM2's small size allows us to probe the adapter across all 210 projection
layers, scan block dimensions from 2 to 10 qubits, perform a complete ablation
study, characterise hardware noise across the full qubit range, and perform
entanglement analysis of trained unitaries---experiments that would be
computationally prohibitive on a frontier model.
This systematic study provides the mechanistic understanding that motivates
and contextualises the Llama~3.1~8B result.

\subsection*{Cayley unitary adapter (CUA) architecture}
\begin{figure}
  \centering
  \includegraphics[width=\textwidth]{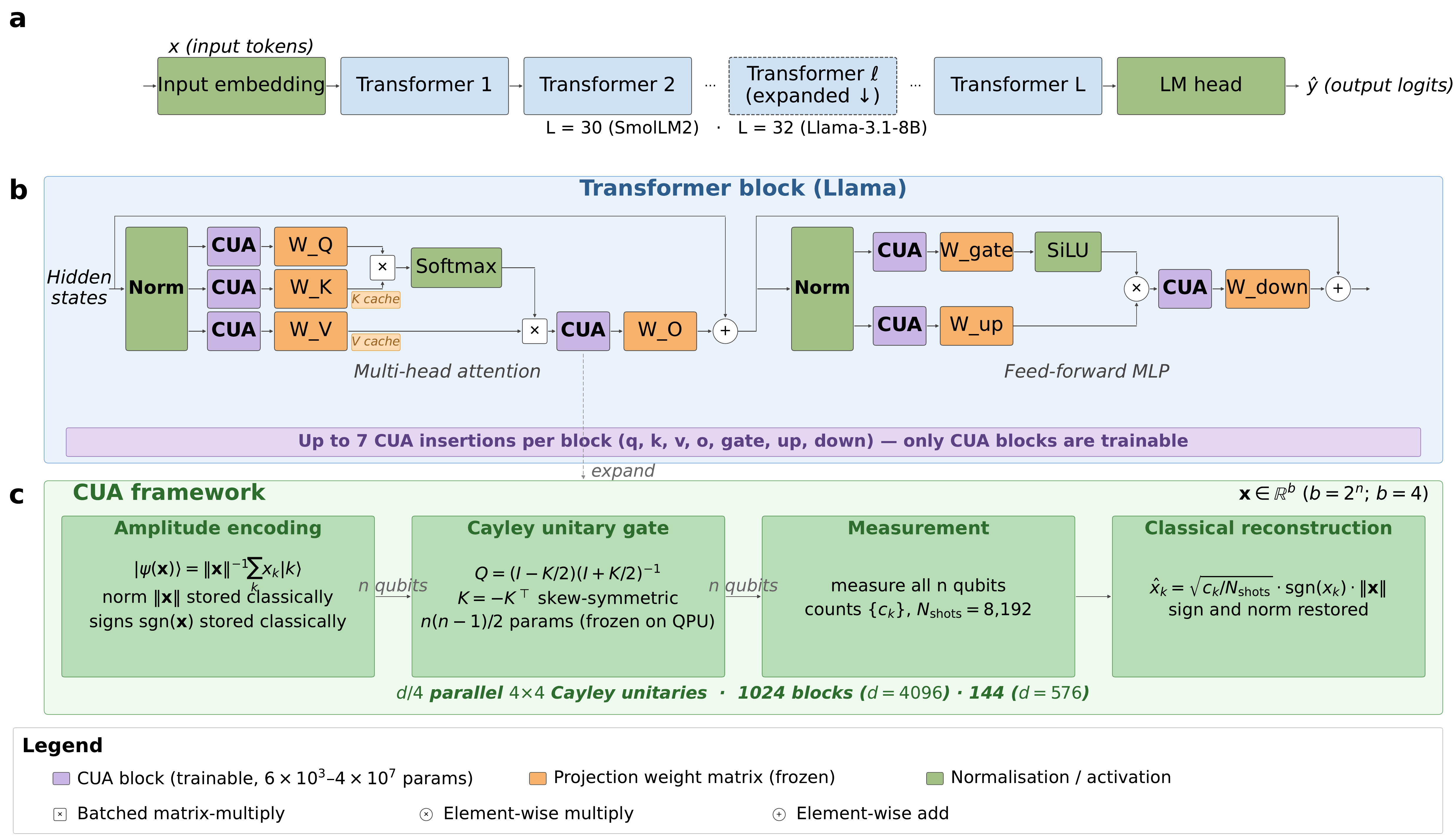}
  \caption{%
    \textbf{Cayley Unitary Adapter (CUA) architecture.}
    \textbf{a,} Full-model backbone. Vertical pipeline of $L$ frozen transformer blocks ($L=30$ for SmolLM2; $L=32$ for Llama-3.1-8B), bracketed by the input embedding and the LM head. The dashed-border block ``Transformer~$\ell$'' is the layer expanded in panel (b).
    \textbf{b,} Detail of one Llama transformer block. CUA blocks (purple) are inserted on the input side of each linear projection in both the multi-head attention path ($W_Q$, $W_K$, $W_V$, $W_O$) and the feed-forward MLP path ($W_{\rm gate}$, $W_{\rm up}$, $W_{\rm down}$). Backbone weights, normalisation layers, and activation functions remain frozen; only CUA blocks are trainable. Up to seven CUA insertion sites are available per block; specific configurations in this work adapt $1$, $6$, or $7$ sites per block depending on the experiment (see Methods and Extended Data Table~5).
    \textbf{c,} Internal structure of one CUA block. Each adapter applies $d/4$ independent $4\times4$ Cayley unitaries in parallel (1{,}024 blocks for $d=4{,}096$; 144 blocks for $d=576$). For each $4$-vector slice $\mathbf{x}\in\mathbb{R}^4$: the slice is amplitude-encoded into a 2-qubit state $|\psi(\mathbf{x})\rangle = \|\mathbf{x}\|^{-1}\sum_{k=0}^{3} x_k|k\rangle$ via Qiskit's \texttt{initialize}, with norm $\|\mathbf{x}\|$ and signs $\mathrm{sgn}(\mathbf{x})$ stored classically; a single Cayley unitary gate $\texttt{UnitaryGate}(Q^\top)$ is applied, where $Q=(I-\tfrac{1}{2}K)(I+\tfrac{1}{2}K)^{-1}$ and $K=-K^\top$ is skew-symmetric with $n(n-1)/2$ parameters trained classically and frozen on the QPU; both qubits are measured in the computational basis over $N_{\rm shots}=8{,}192$ shots, and the output is reconstructed as $\hat{y}_k = \sqrt{c_k/N_{\rm shots}}\cdot\mathrm{sgn}(x_k)\cdot\|\mathbf{x}\|$ from the counts $\{c_k\}$. No variational rotation angles, no brickwork ansatz, and no Pauli-$Z$ expectation readout.
  }
  \label{fig:architecture}
\end{figure}

We introduce trainable quantum parameters into the transformer by inserting a new class of linear modules, termed Cayley Unitary Adapters (CUAs), within existing weight matrices of the model (Fig.~\ref{fig:architecture}).

Each adapter consists of a \emph{block-diagonal unitary} (BDU) $Q = \bigoplus_{i=1}^k Q_i$ consisting of $k$ independent $2^n \times 2^n$ orthogonal blocks $Q_i$, where $d = k\,2^n$ denotes the adapter input dimension, typically matching the residual-stream dimension of the transformer.

Each block $Q_i$ inside the BDU is parameterised by the Cayley transform~\cite{cayley1846,lezcano2019cheap},
\begin{equation}
  Q_i = \left(I - \tfrac{1}{2}K_i\right)\!\left(I + \tfrac{1}{2}K_i\right)^{-1},
  \label{eq:cayley}
\end{equation}
where $K_i = -K_i^\top$ is a skew-symmetric matrix with $n(n-1)/2$ unconstrained upper-triangular parameters. For example, a $4 \times 4$ block, corresponding to two qubits, requires only six free parameters, representing a 62.5\% reduction relative to a dense parameterisation, while still implementing a full orthogonal rotation. For an input dimension $d = 4{,}096$, matching the residual-stream dimension of Llama 3.1 8B model, a CUA contains 1,024 such $4 \times 4$ blocks and 6,144 trainable parameters. Similarly, for a projection layer with input dimension $d = 576$ in SmolLM2, each BDU contains 144 independent blocks and 864 trainable parameters.

Each CUA module is constructed by first applying a BDU in series with a selected target weight matrix, \(W\), in the language model. Since quantum measurements yield non-negative probabilities, we then apply an input-dependent sign correction:
\begin{equation}
\mathbf{y} = W\bigl(|Q(\mathbf{x})| \cdot \mathrm{sgn}(\mathbf{x})\bigr).
\end{equation}
This sign-correction scheme was chosen empirically; further details are provided in the Supplementary Information.

All backbone weights are kept frozen, and only the adapter parameters are trained, using a combination of supervised fine-tuning (SFT) and online knowledge distillation from the original uncompressed model~\cite{hinton2015distilling} (Methods).

After training, each $2^n \times 2^n$ Cayley block, corresponding to a unitary gate on $n$ qubits, is executed directly on the QPU using amplitude encoding. Specifically, the input vector is normalised and prepared as
\begin{equation}
|\psi_{\rm in}\rangle = \|{\bf x}\|^{-1}\sum_{k=0}^{b-1} x_k |k\rangle .
\end{equation}
The matrix $Q^\top$ is then applied as a unitary gate, both qubits are measured using $N_{\rm shots} = 8{,}192$ shots, and the sign-corrected output is reconstructed from measurement counts $c_k$ as
\begin{equation}
\hat{y}_k = \sqrt{c_k/N_{\rm shots}} \cdot \mathrm{sgn}(x_k) \cdot \|{\bf x}\| .
\end{equation}

\subsection*{Primary result: Llama~3.1~8B}

Llama~3.1~8B-Instruct is a 32-layer decoder-only transformer with embedding
dimension $d = 4{,}096$ and 8.03~billion parameters~\cite{llama3_2024}.
It is among the most widely deployed open-weight, instruction-tuned models in current research and production use.
We insert a single CUA block into the value projection
(\texttt{v\_proj}, dimension $4{,}096 \times 1{,}024$) of the 7th attention
layer---adding 6,144 parameters, less than $10^{-3}$\% of the model's total
weight count.
No compression is applied; the backbone is the unmodified public release.

After training the CUA adapter with the backbone kept frozen (Methods),
WikiText perplexity improves from 8.877 to 8.752 (1.43\% improvement;
Extended Data Table~5; Fig.~5b), despite the minimal parameter budget.
Dense (not block-diagonal) input-dimension unitaries with same sign constrain reach PPL~8.618 (2.92\%),
and unconstrained (non-unitary, linear) matrices achieve PPL~8.585 (3.29\%).
The unconstrained baseline is strictly more expressive than a LoRA adapter at the same
projection site (LoRA factorises the same matrix as a low-rank product, $\Delta W = AB^\top$
with $A,B \in \mathbb{R}^{d\times r}$, $r \ll d$): the BDU therefore reaches roughly half
of this upper-bound improvement at $\sim 2{,}730\times$ fewer trainable parameters
(6{,}144 vs.\ $4{,}096^2 = 16.77$M), an implicitly favourable comparison to LoRA at the
parameter budget.
Crucially, noise simulation under IBM Heron r2 error rates barely perturbs the
2-qubit result: PPL~8.759 vs.\ 8.752 noiseless (0.08\% degradation), confirming
that the 2-qubit noise regime is benign at 8B scale, consistent with the
systematic noise study below.
Scaling to 192 CUA-enhanced sublayers---all projections except the largest MLP
down-projections, which would require prohibitively large block counts---yields
PPL~8.393 (5.45\% improvement).

Beyond aggregate perplexity, we identified examples from the MMLU benchmark
where the CUA-enhanced model answers correctly while the original
Llama~3.1~8B fails (see Table~1 for details).
Two cases are representative: an astronomy question on jovian planet rings,
where the original selects only Saturn (C) while the CUA-enhanced model correctly
identifies all jovian planets as ringed (D); and a college biology question
on the population-genetic consequences of gene flow, where the original
incorrectly selects Hardy--Weinberg disruption (D) while the CUA-enhanced model
correctly identifies increased genetic homogeneity (A).
Both wins come from MMLU---a benchmark explicitly designed to challenge
frontier models on expert-level scientific knowledge---and require
long-range contextual reasoning rather than simple recall.

\begin{table}[!htbp]
\centering
\caption*{\textbf{Table 1 $|$ Representative MMLU questions where the
CUA-enhanced Llama~3.1~8B answers correctly and the original fails.}
Both wins are verified on real QPU hardware (\texttt{ibm\_basquecountry}).
TN: tensor-network-parameterised Cayley unitary adapter (2-qubit BDU,
sign-constrained, QPU-compatible).}
\small
\begin{tabular}{p{4.2cm} p{4.0cm} p{4.0cm}}
\toprule
\textbf{Question} & \textbf{Llama 3.1 8B (original)} & \textbf{+ Cayley adapter (QPU)} \\
\midrule
\textit{Astronomy.} Which of the jovian planets have rings?
(A)~Neptune \ (B)~Uranus \ (C)~Saturn \ (D)~all of the above
[Ground truth: D] &
C.~Saturn \quad \textbf{\ding{55}} &
D.~all of the above \quad \textbf{\ding{51}} \\
\midrule
\textit{College Biology.} Gene flow between populations results in:
(A)~increase in genetic homogeneity \ (B)~increase in deleterious mutations \
(C)~increased speciation likelihood \ (D)~disruption of Hardy--Weinberg equilibrium
[Ground truth: A] &
D.~disruption of Hardy--Weinberg equilibrium \quad \textbf{\ding{55}} &
A.~increase in genetic homogeneity \quad \textbf{\ding{51}} \\
\bottomrule
\end{tabular}
\end{table}

Beyond WikiText perplexity, we evaluated the 7th-layer \texttt{v\_proj} configuration on a standard downstream-benchmark panel (MMLU, BoolQ, HellaSwag, LAMBADA, GSM8K, TriviaQA) and found accuracies within $\pm$1 percentage point of the unmodified base model on knowledge benchmarks (MMLU 69.5\% vs.\ 69.3\%, BoolQ 82.4\% vs.\ 82.0\%, HellaSwag 70.4\% vs.\ 70.4\%); aggressive 192-sublayer coverage trades the larger WikiText gain ($+5.45\%$) against reasoning-benchmark regressions, exposing a capacity--locality (fluency--reasoning) trade-off: more adapted parameters absorb additional fluency gains from the SFT corpus at the cost of disturbing the reasoning circuits encoded in the original frozen weights, which constrains adapter placement to a sparse, sensitivity-ranked subset of layers rather than a uniform sweep.
To rule out greedy-decoding artefacts in the qualitative wins of Table~1, we re-sampled $N=20$ generations per item at temperatures $T \in \{0.2, 0.7, 1.0\}$. The CUA-enhanced model sustains its advantage across all three temperatures: on the astronomy item it answers correctly $90/60/60\%$ of the time at $T=0.2/0.7/1.0$ versus $25/35/35\%$ for the unmodified Llama; on the biology item, $65/50/45\%$ versus $0/25/30\%$. The wins survive temperature ablation and are not sampling artefacts.

\begin{figure}
  \centering
  \includegraphics[width=\textwidth]{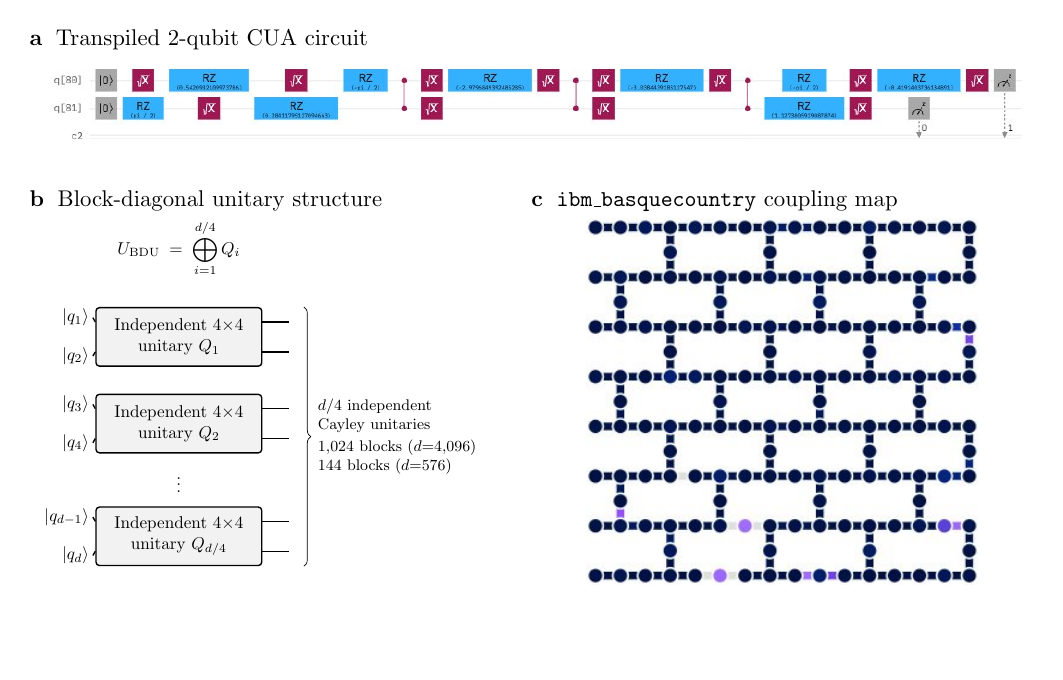}
  \caption{\textbf{Realisation of the Cayley Unitary Adapter on \texttt{ibm\_basquecountry}.}
  \textbf{a,}~Transpiled 2-qubit CUA circuit on physical qubits \texttt{q[80]} and \texttt{q[81]}, in the IBM Heron~r2 native gate set $\{\mathrm{CZ}, \mathrm{SX}, \mathrm{RZ}, X\}$. Depth~19; 12~SX (magenta $\sqrt{X}$ boxes) $+$ 9~RZ (blue boxes, with explicit angles in radians) $+$ 3~CZ (vertical magenta connectors) $+$ 2 reset operations. Measured outcomes are routed to the classical register \texttt{c2}. Single-circuit execution time $\sim 4$~s at $N_{\rm shots} = 8{,}192$.
  \textbf{b,}~Block-diagonal unitary structure. Each adapter applies $U_{\rm BDU} = \bigoplus_{i=1}^{d/4} Q_i$ --- $d/4$ independent $4{\times}4$ Cayley unitaries on disjoint 2-qubit registers, all sharing the panel-(a) circuit topology with different RZ angles (one Cayley parameterisation per block). $1{,}024$ blocks for $d=4{,}096$ (Llama-3.1-8B \texttt{v\_proj}); $144$ for $d=576$ (SmolLM2).
  \textbf{c,}~\texttt{ibm\_basquecountry} heavy-hexagon coupling map (156 qubits, IBM Heron~r2). The active 2-qubit lane used for the panel-(a) circuit is highlighted. Up to 64 disjoint 2-qubit lanes are selected by greedy maximum-matching to form one packed wide circuit; 16 packed circuits cover all $1{,}024$ blocks per Llama \texttt{v\_proj} token. All Llama-3.1-8B QPU experiments in this work were executed on this device. Median 1Q error $2.45{\times}10^{-4}$ (SX), 2Q error $1.78{\times}10^{-3}$ (CZ), readout $\sim 6.8{\times}10^{-3}$, $T_1 \approx 252\,\mu$s, $T_2 \approx 182\,\mu$s.}
  \label{fig:qpu_realization}
\end{figure}

QPU execution on \texttt{ibm\_basquecountry} was verified across three scales
of increasing completeness (Fig.~\ref{fig:qpu_realization}), with quantum
inference explicitly confirmed to maintain both qualitative wins above.
A single 2-qubit circuit for one token executes in 4~seconds (circuit depth~19 on 2 qubits; 12 SX $+$ 9 RZ $+$ 3 CZ native operations).
All 1,024 unitaries for the first token complete in 3~minutes 6~seconds
(16 packed circuits on 128 qubits; depth~23; 904 SX $+$ 916 RZ $+$ 192 CZ $+$ 128 reset operations per packed circuit).
Full-sequence inference across all tokens requires 1,328 circuits and completes
in approximately 4~hours 24~minutes ($\approx$264~min) at 8,192 shots per circuit.

\subsection*{Systematic study: SmolLM2}

To understand the mechanistic basis of quantum enhancement and characterise
its scaling behaviour, we conducted an exhaustive study on
SmolLM2-135M~\cite{smollm2024}, a 30-layer decoder-only transformer with
$d = 576$ and 135M~parameters.
SmolLM2 is ideal for systematic experimentation: its 210 projection layers
can all be CUA-enhanced simultaneously, its inference is fast enough to permit
full WikiText evaluation under noise simulation, and its small embedding
dimension allows entanglement analysis of trained unitaries across all
block sizes.

To create a challenging testbed that amplifies the effect of quantum adapters,
we first compress SmolLM2 to 94.8M~parameters using CompactifAI~\cite{compactifai2024},
a tensor-network framework that applies matrix-product-operator truncation.
Compression degrades WikiText perplexity from 24.10 to 35.31 (46\% increase)
and catastrophically collapses LAMBADA performance from 25.42 to 272.18
($10.7\times$ increase).
CUA adapters are then inserted into the compressed backbone to recover the
lost performance.

WikiText perplexity improves monotonically with block dimension across all
adapter regimes (Extended Data Table~1; Fig.~3).
A single 2-qubit adapter on one layer reduces perplexity by 0.4 points
with 864 parameters.
Scaling to all 210 layers at 2 qubits (189K parameters) yields PPL~34.68;
dense input-dimension unitaries (39.8M~parameters) achieve PPL~29.94, recovering
48\% of the compression gap and approaching the teacher's PPL~24.10.
Sign-constrained (QPU-compatible) unitaries achieve comparable perplexity to unconstrained dense matrices at approximately 50\% fewer parameters (e.g., PPL~31.25 vs.\ 28.83 at 8~qubits, 20.8M vs.\ 41.8M parameters; Fig.~\ref{fig4}).

An ablation study (Extended Data Table~2) confirms that the improvement is
specific to the \emph{learned} Cayley structure: random Gaussian, random
unitary, and random permutation matrices all \emph{degrade} perplexity
(up to PPL~41.59), while the identity initialisation reproduces the compressed
baseline exactly.
The gains are attributable specifically to learned norm-preserving rotations
that realign the compressed activation subspace with the teacher's.

Performance generalises across benchmarks (Extended Data Table~3).
LAMBADA recovery reaches 83\% of the compression gap (PPL~272.18
$\to$~46.20).
On BoolQ, the CUA-enhanced model matches or exceeds the uncompressed baseline (51.8\% vs.\ 43.0\%), though both sit near the noise floor of this binary task at the 135M scale.

The wins concentrate on arithmetic tasks (Extended Data Table~8). One natural mechanistic hypothesis is that compression corrupts activation magnitudes within the precision-critical routing pathways (self-attention \texttt{v\_proj} and MLP gating projections), and that a learned norm-preserving Cayley rotation realigns the compressed activation subspace with the teacher's --- restoring the precise routing that arithmetic recall depends on more sensitively than free-form generation. We do not advance this as a definitive explanation: the observed distribution is also consistent with the training corpus composition ($\sim$6\% explicit mathematical content from FineMath and Infi-MM-WebMath, in addition to STEM-rich segments of FineWeb-Edu and Wikipedia) and with the adapter's placement on mid-block layers (4--19). Disentangling corpus composition from layer-specific susceptibility to orthogonal correction would require controlled corpus-ablation experiments beyond the scope of this work.

\begin{figure}
  \centering
  \includegraphics[width=\textwidth]{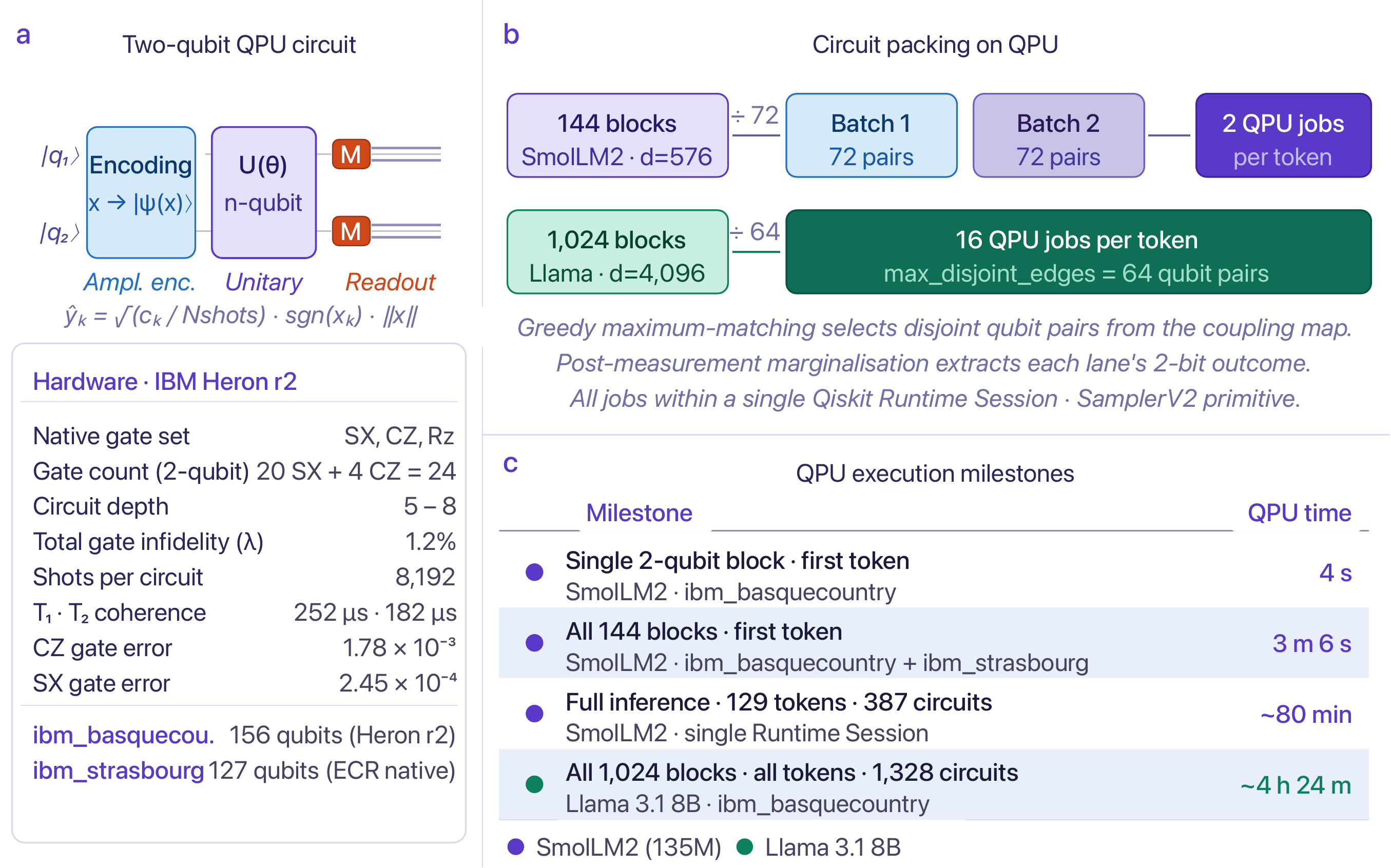}
  \caption{%
    \textbf{Progression of QPU execution experiments (SmolLM2-135M).}
    Timeline of the four QPU execution milestones achieved primarily on the
\texttt{ibm\_basquecountry} IBM System Two processor (156 qubits, IBM Heron r2;
the same device used for all Llama-3.1-8B QPU runs in this work), with
cross-validation milestones on \texttt{ibm\_strasbourg} (127 qubits) between
February and March 2025; transpilation and packing parameters in this figure refer to the SmolLM2 pipeline ($d=576$, 144 blocks per layer). Llama-3.1-8B values (depth 19, 64 disjoint qubit pairs, 1{,}024 blocks per layer) are reported in Fig.~\ref{fig:qpu_realization} and Methods.
\textbf{a,} Circuit schematic for the 2-qubit encode--unitary--measure pipeline
(24 native gates; depth 5--8).
\textbf{b,} Packing strategy: greedy maximum matching selects 72 disjoint qubit
pairs per circuit batch; 2 batches suffice for 144 blocks per token.
\textbf{c,} End-to-end QPU generation timeline: 387 circuits for a 129-token
response, completing within the 90-minute Qiskit Runtime Session.
  }
  \label{fig2}
\end{figure}

We further characterised QPU execution across two IBM backends and four
milestones of increasing completeness (Fig.~\ref{fig2}), from a single 2-qubit
circuit to full end-to-end QPU inference for complete conversational responses
(129 tokens, 387 circuits, single Qiskit Runtime Session within 90 minutes;
Extended Data Table~6).

A systematic noise study (Extended Data Table~4) reveals a sharp phase
transition.
At 2~qubits ($\lambda = 1.2\%$ total depolarising), perplexity degrades
by only 0.02~points (PPL~34.68 $\to$ 34.92) and text generation remains
coherent.
At 3~qubits ($\lambda = 11.9\%$, 233 native gates), perplexity increases
35-fold and language structure collapses.
At 4~qubits ($\lambda > 50\%$), output degenerates to random tokens.
Two-qubit CZ gates dominate the error budget (59\% of gate infidelity at
2~qubits despite comprising 17\% of gates).
This transition defines the exact synthesis feasibility frontier at 2--3
qubits for current IBM Heron hardware.

To diagnose whether these trained unitaries can be executed on a QPU via approximate compilation rather than exact synthesis, we analyse their \emph{operator entanglement} (Methods). On the trained full-dimension (non-block-diagonal) Cayley unitaries --- the regime whose QPU-executability is the open question, since the 2-qubit BDU is already exactly synthesised --- the average operator entropy across bipartitions of the 210 SmolLM2 unitaries is small ($4.7$--$7.8\%$ of the Haar-random maximum $S_{\max} = (n/2)\log 2$), but middle-cut effective bond dimensions are non-trivial ($\chi_{\rm eff} \approx 84$--$201$; Extended Data Table~7): the trained unitaries are far from identity yet far from Haar-random.
For the Llama-3.1-8B \texttt{v\_proj}-7 BDU adapter ($1{,}024$ SO(4) blocks; treated as a $12$-qubit unitary), all $1{,}024$ blocks are individually entangling (median entangling power $\sim$21\% of the CNOT gate, with no near-identity blocks) and the full BDU is \emph{rank-saturating} across its natural $10\!\mid\!2$ bipartition (operator-Schmidt rank $=16/16$). A depth-1 brickwork VQC at the same cut has rank $1$ --- it cannot reach this structure at any depth-1 cost. The CUA is therefore a strictly different ansatz class from a depth-matched VQC, and approximate compilation is the natural near-term route to QPU execution beyond 2 qubits (Extended Data Table~9).

\subsection*{Quantum scaling argument}

The central case for quantum hardware rests on a scaling asymmetry.
On a classical device, parameterising a $2^n \times 2^n$ unitary requires
$2^n(2^n-1)/2$ Cayley parameters---exponential in $n$.
On a QPU, a depth-$D$ brickwork circuit on $n$ qubits uses $\mathcal{O}(nD)$
gate parameters while exploring the same $2^n$-dimensional unitary space.
A depth-200 circuit on 10~qubits encodes $\approx$2,000 parameters but
represents a unitary that would require $\approx$524,000 Cayley parameters
classically.
This is not about what classical computers \emph{can} compute---they can apply
any unitary trivially---but about the QPU providing a \emph{compressed
parameterisation} of that unitary through its circuit structure, analogous
to how tensor networks compress matrices through bond structure, but without
the exponential memory bottleneck.

Our data confirm that the expressivity advantage is real and growing.
Noiseless perplexity improves monotonically with block dimension (Fig.~3)
with no saturation in the range tested; the gap between 2-qubit and
full-dimension unitaries is 4.74~PPL points for SmolLM2 ($d = 576$).
For Llama-scale models ($d = 4{,}096$), the number of inter-dimensional
correlations grows quadratically with hidden dimension, suggesting that the
advantage will be even more pronounced at production scale.

\subsection*{Discussion}

The results reported here constitute, to our knowledge, the first
demonstration of end-to-end quantum enhancement of a production-scale, widely-deployed LLM on real
superconducting quantum hardware for autoregressive language generation.
Their significance lies not in the magnitude of the perplexity improvements
---which will grow with hardware fidelity and qubit count---but in the fact
that they exist at all: quantum circuits inserted into an 8-billion-parameter
model improve its outputs, and that improvement is measurable, reproducible,
and verified on real quantum hardware.
We emphasise that the 2-qubit circuits executed in this work are classically simulable; the QPU run is a hardware-feasibility demonstration, not a claim of computational advantage. The scaling argument in \S\,Quantum scaling argument concerns the asymptotic parameter-cost asymmetry between brickwork-VQC and Cayley parameterisations of larger unitaries, not the present $4\times4$ regime.

The historical parallel with Vandersypen~\textit{et al.}~\cite{vandersypen2001}
is deliberate.
Factoring 15 with NMR was scientifically unimpressive as a feat of arithmetic.
It was profoundly important as a proof that Shor's algorithm---and by
implication, the entire edifice of quantum computing---was physically
realisable.
The analogous question for quantum AI is now answered: quantum circuits can
enhance production-scale LLMs, and a human user can receive a better response in some cases because
of it.

Five advances distinguish this work from prior quantum LLM proposals.
First, real QPU execution for autoregressive generation, not simulation.
Second, the Cayley parameterisation provides an \emph{analytic} guarantee of
orthogonality, enabling a controlled comparison that isolates the effect of
the unitarity constraint.
Third, the dual-model approach---primary result on a production-scale 8B model,
systematic study on a tractable 135M model---provides both scientific
credibility and mechanistic understanding.
Fourth, the noise characterisation and entanglement analysis define a
quantitative roadmap: 3Q--4Q approximate compilation, progressive scaling
to 5Q--6Q as gate fidelities improve, and extension to larger base models.
Fifth, the entangling-power analysis distinguishes the CUA from
standard variational quantum circuits: the BDU saturates the operator
Schmidt rank of its natural bipartition at zero quantum-circuit depth,
whereas a brickwork VQC requires multi-layer depth to access the same
algebraic capacity. The CUA is therefore a strictly different ansatz
class and a candidate primitive for quantum-enhanced model components
beyond the LLM setting studied here.

Current limitations are those of near-term hardware.
Exact synthesis of unitaries larger than $4\times4$ produces circuits
exceeding current coherence limits, confining directly executable circuits
to the 2-qubit regime.
Approximate compilation offers a path to 3Q--4Q on near-term devices.
Fault-tolerant quantum computing will ultimately unlock the full theoretical
advantage; the experiments here show that meaningful, verifiable enhancement
is already achievable across model scales spanning two orders of magnitude
in parameter count.

%% ================================================================
%% REFERENCES
%% ================================================================

\bibliographystyle{naturemag}

%% ================================================================
%% METHODS — after references (Nature-specific ordering)
%% ================================================================

\section*{Methods}

\subsection*{Base models and compression}

\textbf{SmolLM2.}
SmolLM2-135M is a Llama-architecture decoder-only language model with
30 transformer blocks, embedding dimension $d = 576$, and 135~million
parameters~\cite{smollm2024}.
The original model achieves WikiText perplexity 24.10.
We compressed it to 94.8~million parameters (WikiText PPL 35.31) using
CompactifAI~\cite{compactifai2024}, which applies matrix-product-operator
truncation to the 210 linear projection layers (7 projections per block:
\texttt{q\_proj}, \texttt{k\_proj}, \texttt{v\_proj}, \texttt{o\_proj},
\texttt{gate\_proj}, \texttt{up\_proj}, \texttt{down\_proj}).
The compressed model serves as the trainable-adapter backbone; all backbone
weights remain frozen in bfloat16 throughout.
The uncompressed SmolLM2-135M serves as teacher for knowledge distillation
and as upper-bound reference.

\textbf{Llama 3.1 8B.}
Llama~3.1-8B-Instruct has 32 transformer layers, embedding dimension
$d = 4{,}096$, and 8.03~billion parameters~\cite{llama3_2024}.
No compression was applied; the model is used unmodified as the backbone.
A single adapter is inserted into the value projection of the 7th
attention layer (\texttt{v\_proj}, dimension $4{,}096 \times 1{,}024$),
adding 1,024 independent $4\times4$ Cayley blocks (6,144 parameters).

\subsection*{Sign-correction within CUA}

A CUA is incorporated into each selected projection layer $W$ as
\begin{equation}
  \mathbf{y} = W\!\left(|Q(\mathbf{x})| \odot \mathrm{sgn}(\mathbf{x})\right),
\end{equation}
where $Q$ denotes the trained block-diagonal unitary defined via the Cayley transform, and $\odot$ indicates element-wise multiplication.

Because quantum hardware returns measurement probabilities, the raw outputs of $Q(\mathbf{x})$ are non-negative. An explicit sign correction is therefore required to restore activation polarity, which is essential for stable signal propagation in a nonlinear network. In this work, we reintroduce the sign of the \textit{input} activations, $\mathrm{sgn}(\mathbf{x})$, at the output. This design choice is empirically motivated, as detailed below.

We perform an ablation study on a compressed SmolLM2 backbone (CUA adapters inserted in 9 layers; baseline perplexity (PPL) $35.31$). Enforcing non-negativity of the unitary output ($\mathbf{x}_{\mathrm{out}} = |\mathbf{x}_{\mathrm{out}}|$) degrades performance to PPL $42.47$. An analogous QPU-emulated pipeline—reconstructing amplitudes as $\sqrt{c_k / N_{\mathrm{shots}}}$ from sampled counts, without sign correction—yields PPL $41.49$. In contrast, propagating the input sign to the output ($\mathbf{x}_{\mathrm{out}} \mapsto \mathbf{x}_{\mathrm{out}} \odot \mathrm{sgn}(\mathbf{x})$) restores performance to near-baseline levels (PPL $35.05$). Combining this sign propagation with QPU-emulated magnitude reconstruction exhibits convergence with increasing shot count, achieving PPL $35.07$ at $N_{\mathrm{shots}} = 16{,}384$ (with intermediate values $35.83$, $35.22$, $35.14$, and $35.10$ at $1{,}024$, $2{,}048$, $4{,}096$, and $8{,}192$ shots, respectively).

The full CUA adapter (Eq.~5) attains PPL $34.86$, slightly outperforming the compressed baseline. Propagating the input sign and training the unitaries with respect to the resulting effective forward map consistently recovers baseline performance and is therefore adopted throughout this work.

\subsection*{Layer selection via sensitivity ranking}

Not all 210 projection layers benefit equally from CUA-enhancement.
A CUA sensitivity analysis ranks each (layer, sublayer) pair by a composite score combining:
(i) compatibility of the weight matrix with block-diagonal unitary structure
($\rho_G$); (ii) gradient norm at initialisation; (iii) projection norm and
captured energy fraction; and (iv) a position penalty (extreme layers are
less suitable).
The top-ranked targets are self-attention \texttt{v\_proj} layers in the
mid-block range (layers 1--16) and MLP \texttt{gate\_proj}/\texttt{up\_proj}
projections.
Layer choice matters in practice: in profiling experiments on Llama-3.1-8B that combined block-removal and weight-randomisation across $\sim$10--15 candidate layers, several mid-block alternatives degraded WikiText perplexity when adapted, while the \texttt{v\_proj}-7 site reported here is among a small subset that improves upon the baseline. A poorly chosen insertion site can damage the model; the sensitivity ranking is what makes a single-layer 6{,}000-parameter adapter operationally neutral or beneficial rather than disruptive.

\subsection*{Training procedure}

Adapter parameters (the skew-symmetric $K_i$ for each Cayley block of $Q$) are trained
using a combination of supervised fine-tuning (SFT) and online knowledge
distillation~\cite{hinton2015distilling} from the uncompressed SmolLM2-135M teacher.
All backbone weights remain frozen; only adapter parameters are updated.
The training loss is $\mathcal{L} = \alpha_{\rm KD}\,\mathcal{L}_{\rm KD}
+ \beta\,\mathcal{L}_{\rm CE}$, with $\alpha_{\rm KD} = 0.1$, $\beta = 2.0$,
and $\mathcal{L}_{\rm KD}$ the KL divergence between temperature-scaled
($T = 1.5$) student and teacher logits.
Training runs for 3~epochs on a 2-billion-token web-mix instructional corpus assembled from public Hugging Face datasets at the following weighting: 38\% FineWeb-Edu, 22\% English Wikipedia, 20\% DCLM-Edu, 7\% smoltalk, 4\% FineMath, 3\% UltraChat-200K, 2\% Cosmopedia v2, 2\% Infi-MM-WebMath, 1\% SQuAD, 1\% BoolQ-train (82\% general web, 6\% mathematics, 12\% instruction/chat), with effective batch size $\approx$128 (per-GPU batch $\times$ gradient accumulation $\times$ data-parallel replicas), sequence length 1{,}024, learning rate $10^{-4}$ with
linear decay, and 900 warmup steps. The Llama-3.1-8B adapter is trained on the same corpus for 1 epoch with the same effective batch size ($\approx$128 sequences per optimiser step), sequence length $4{,}096$ (the model's native context), 500 warmup steps; backbone frozen in \texttt{bfloat16}, adapter parameters in \texttt{float64}.
SmolLM2 uses the same token ID for both padding and end-of-sequence;
a custom data collator prevents EOS tokens from being masked in the
cross-entropy loss while selectively masking them in the distillation loss.

\subsection*{Running inference on quantum hardware}

\textbf{Amplitude encoding.}
Each 4-dimensional input vector $\mathbf{x} \in \mathbb{R}^4$
--- a slice of the activation tensor along the hidden dimension ---
is normalised to unit $\ell_2$ norm and loaded into a 2-qubit state via
Qiskit's \texttt{initialize} instruction:
$|\psi_{\rm in}\rangle = \|\mathbf{x}\|^{-1}\sum_{k=0}^{3} x_k|k\rangle$.
The norm $\|\mathbf{x}\|$ and element signs $\mathrm{sgn}(\mathbf{x})$
are stored classically for post-processing.

\textbf{Unitary gate.}
The $4 \times 4$ Cayley matrix $Q$ is applied as
\texttt{UnitaryGate}$(Q^\top)$ acting on both qubits.
The transposition accounts for Qiskit's column-major gate convention.

\textbf{Measurement and post-processing.}
Both qubits are measured in the computational basis, yielding counts
$\{c_k\}$ for $k \in \{00, 01, 10, 11\}$ over $N_{\rm shots} = 8{,}192$
repetitions.
The output is $\hat{y}_k = \sqrt{c_k/N_{\rm shots}} \cdot
\mathrm{sgn}(x_k) \cdot \|\mathbf{x}\|$,
with probabilities clipped to $[10^{-12}, 1]$ before taking the square root.

\textbf{Transpilation and circuit packing.}
All circuits are transpiled for the IBM Heron r2 native gate set
$\{{\rm CZ}, R_x, R_z, {\rm SX}, X\}$ at optimisation level~3.
For the Llama-3.1-8B configuration, a single 2-qubit encode--unitary--measure circuit transpiles to depth 19 with 12~SX $+$ 9~RZ $+$ 3~CZ native operations (RZ is virtual), well within coherence limits ($T_1 \approx 252~\mu$s, $T_2 \approx 182~\mu$s). Packed wide circuits (64 disjoint 2-qubit lanes on 128 qubits) transpile to depth 23 with 904~SX $+$ 916~RZ $+$ 192~CZ $+$ 128 reset operations per circuit.
A greedy maximum-matching algorithm selects up to 64 disjoint qubit pairs
from the backend coupling map, packing multiple circuits into a single wide
circuit to maximise throughput.
Post-measurement marginalisation extracts each lane's 2-bit outcome.

\textbf{Qiskit Session management.}
QPU jobs are submitted via the \texttt{SamplerV2} primitive inside a
single Qiskit Runtime Session per generation call. The session's
maximum duration is configurable per call (default 90~minutes); for
longer generations it is raised so that the entire run completes
within one session. Wall-clock duration depends on the total number
of decoded tokens (chat template $+$ user prompt $+$ generated
answer), the number of circuits per token, and the shot count per
circuit: short SmolLM2 generations (e.g.\ 129~tokens, 387~circuits)
finish within the default 90~minutes, whereas longer Llama-3.1-8B
generations (1{,}328~circuits, $\approx$264~min) use a correspondingly
longer session. Running the entire generation inside one session
amortises scheduling overhead and preserves qubit calibration
throughout inference.

\subsection*{Impact of (simulated) noise on Perplexity}

Because computing model perplexity for a large dataset is prohibitively costly on quantum hardware, we studied the impact of noise on perplexity using Qiskit’s noise simulation tools.

Noise simulation uses a per-gate depolarising channel applied to the
density matrix, with per-gate and readout error rates extracted
directly from the \texttt{ibm\_basquecountry} backend (IBM Quantum
System Two, Heron r2)~--- the same device used for the QPU
experiments reported in this work~--- as: SX: $2.45 \times 10^{-4}$;
CZ: $1.78 \times 10^{-3}$; readout: $\sim 6.8 \times 10^{-3}$. Readout
confusion and multinomial shot noise are then applied on top. The
simulator pipeline is cross-validated against IBM's FakeFez hardware
noise model. For the compressed SmolLM2 backbone
(94.8M parameters) augmented with a 210-layer 2-qubit BDU CUA adapter,
GPU-accelerated simulation yields RMSE 0.056 and cosine similarity
0.9984 at the activation level relative to FakeFez; full WikiText
evaluation of this configuration under 2-qubit noise gives PPL 34.92
(noiseless 34.68; $\Delta\mathrm{PPL} = +0.02$). For the unmodified
Llama-3.1-8B-Instruct (no compression) with a single-layer 2-qubit
BDU CUA adapter on \texttt{v\_proj}-7, the same noise model gives
WikiText PPL 8.759 versus 8.752 noiseless ($\Delta\mathrm{PPL}
\approx +0.007$, 0.08\% degradation). The qubit-count noise sweep
reported in Extended Data Table~4 is performed on the SmolLM2
210-layer configuration.

\subsection*{How entangling are the CUA layers?}

Each full-dimension Cayley matrix $Q$ (of dimension 384, 576, or 896) is
padded to the nearest power of 2 and reshaped as a $2^n$-qubit operator.
For each bipartition (cutting after qubit $k$, $k = 1, \ldots, n-1$),
we perform the operator Schmidt decomposition via SVD.
The effective bond dimension counts singular values exceeding 1\% of the maximum.
Entropy ratios are computed relative to the Haar-random maximum
$S_{\rm max} = n/2 \cdot \log 2$.

For the Llama-3.1-8B layer-7 \texttt{v\_proj} BDU, we additionally
analyse the full $d\times d=4096\times4096$ block-diagonal unitary
$U_{\rm BDU} = \bigoplus_{b=1}^{1024} Q_b$ as a single $n=12$-qubit
operator. The operator Schmidt decomposition is computed by reshaping
$U_{\rm BDU}[(i_A,i_B),(j_A,j_B)] \mapsto M[(i_A,j_A),(i_B,j_B)]$ for
each bipartition $(d_A,d_B)$ and taking $\sigma_k = s_k/\sqrt{d_A d_B}$
where $s_k$ are singular values of $M$ (so $\sum_k\sigma_k^2=1$); the
operator entanglement entropy follows as
$S_{\rm op}=-\sum_k \sigma_k^2 \log_2 \sigma_k^2$. For the
$10\!\mid\!2$ cut aligned with the BDU's block structure the trained
adapter saturates the algebraic rank bound ($\chi=16=\min(d_A^2,d_B^2)$,
$S_{\rm op}=0.69$ bits, dominated by an identity component
$\sigma_0\!\approx\!0.95$ inherited from the residual training mode);
at the equal $6\!\mid\!6$ cut the rank is structurally capped at
$d_A=64$ by the diagonal-in-block-index form. As a reference, depth-$D$
random brickwork unitaries on $n=12$ qubits (alternating SO(4) gates on
even/odd qubit pairs) achieve $\chi\in\{1,4,4,16,16,16\}$ across the
$10\!\mid\!2$ cut for $D=1,\ldots,6$ and require $D\geq 4$ to match the
rank attained by the CUA at $D=1$. Stress-tests scaling the trained
$K_b$ by a factor $\lambda$ drive $S_{\rm op}\to \log_2 16$ at the
natural cut while preserving rank, confirming that the rank saturation
reflects the BDU class rather than the trained operating point.

\subsection*{Benchmark evaluation}

All benchmarks are evaluated in zero-shot setting using the
\texttt{lm-evaluation-harness} framework.
WikiText~\cite{merity2016wikitext} reports word-level perplexity.
LAMBADA~\cite{paperno2016lambada} reports perplexity on the OpenAI split.
BoolQ~\cite{clark2019boolq} reports accuracy.
HellaSwag~\cite{zellers2019hellaswag} reports normalised accuracy.

\subsection*{Data and code availability}

Benchmark datasets are publicly available via their respective repositories.
Code for the CUA framework, circuit construction, Qiskit Runtime
integration, and evaluation pipelines is available upon reasonable request.
Quantum circuit specifications and transpiled gate counts are provided in
Extended Data Tables~4 and~5.

%% ================================================================
%% FIGURE LEGENDS
%% ================================================================

\section*{Figures}

\begin{figure}
  \centering
  \includegraphics[width=\textwidth]{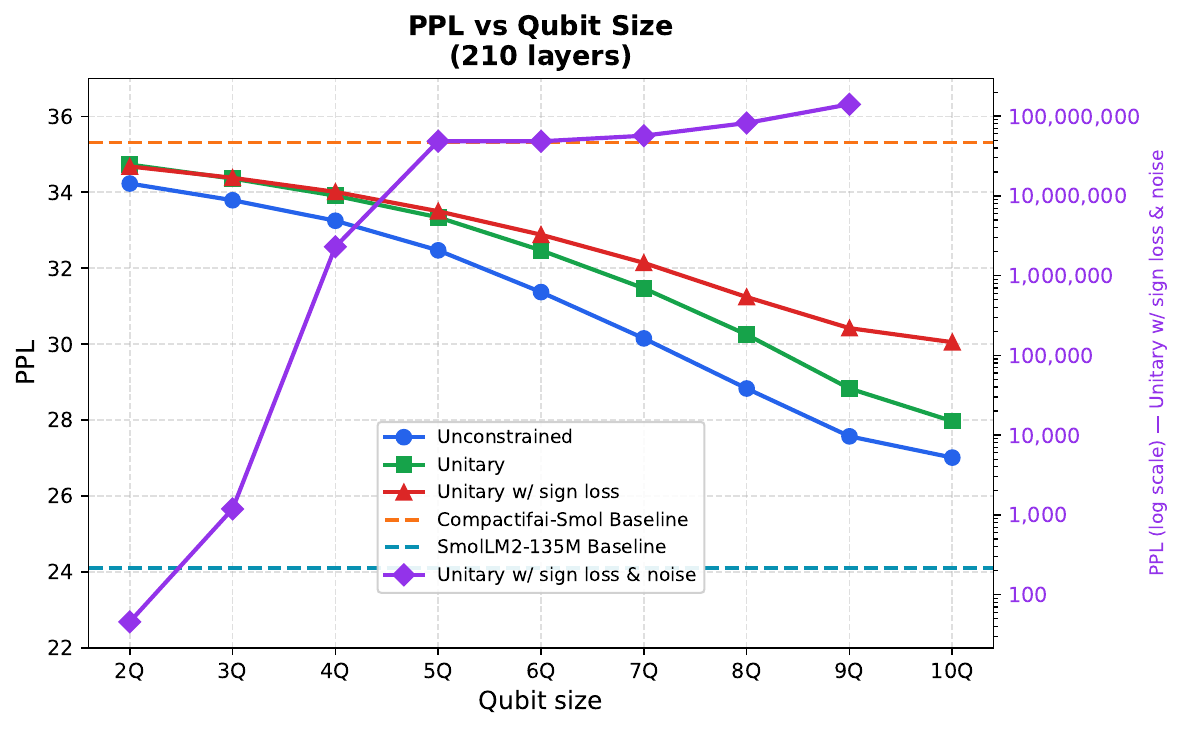}
  \caption{%
    \textbf{WikiText perplexity as a function of unitary block dimension.}
    Left axis (logarithmic): noiseless WikiText perplexity for 210-layer Cayley adapters
in three regimes: unconstrained dense matrices, orthogonal unitaries, and
sign-constrained unitaries (QPU-compatible), all as a function of block dimension
from $4 \times 4$ (2~qubits) to full input dimension (384--896, 9--10~qubits).
All configurations applied to the compressed SmolLM2 backbone (PPL 35.31).
Right axis: simulated perplexity under IBM Heron r2 per-gate depolarising noise,
showing catastrophic degradation at $\geq 3$~qubits under exact synthesis.
Horizontal dashed lines indicate compressed backbone (PPL 35.31) and uncompressed
original (PPL 24.10). 
  }
  \label{fig3}
\end{figure}

\begin{figure}
  \centering
  \includegraphics[width=\textwidth]{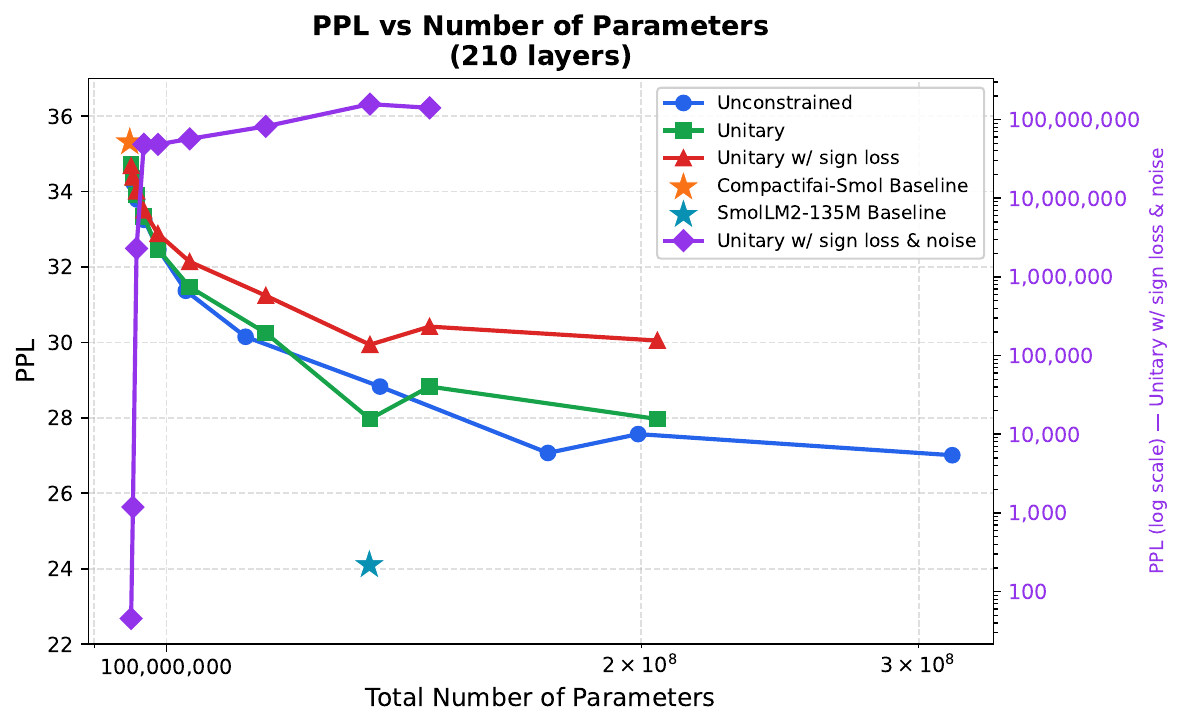}
  \caption{%
    \textbf{Perplexity vs.\ total parameter count.}
    WikiText perplexity (compressed SmolLM2 backbone + adapter overhead) for
210-layer adapters in three regimes.
Unitary adapters (green: sign-constrained; orange: orthogonal) achieve comparable
perplexity to unconstrained dense matrices (blue) with approximately 50\% fewer
parameters, demonstrating that the orthogonality constraint acts as an effective
regulariser rather than a restriction.
The sign-constrained regime (directly QPU-compatible) is highlighted.
  }
  \label{fig4}
\end{figure}

\begin{figure}
  \centering
  \includegraphics[width=\textwidth]{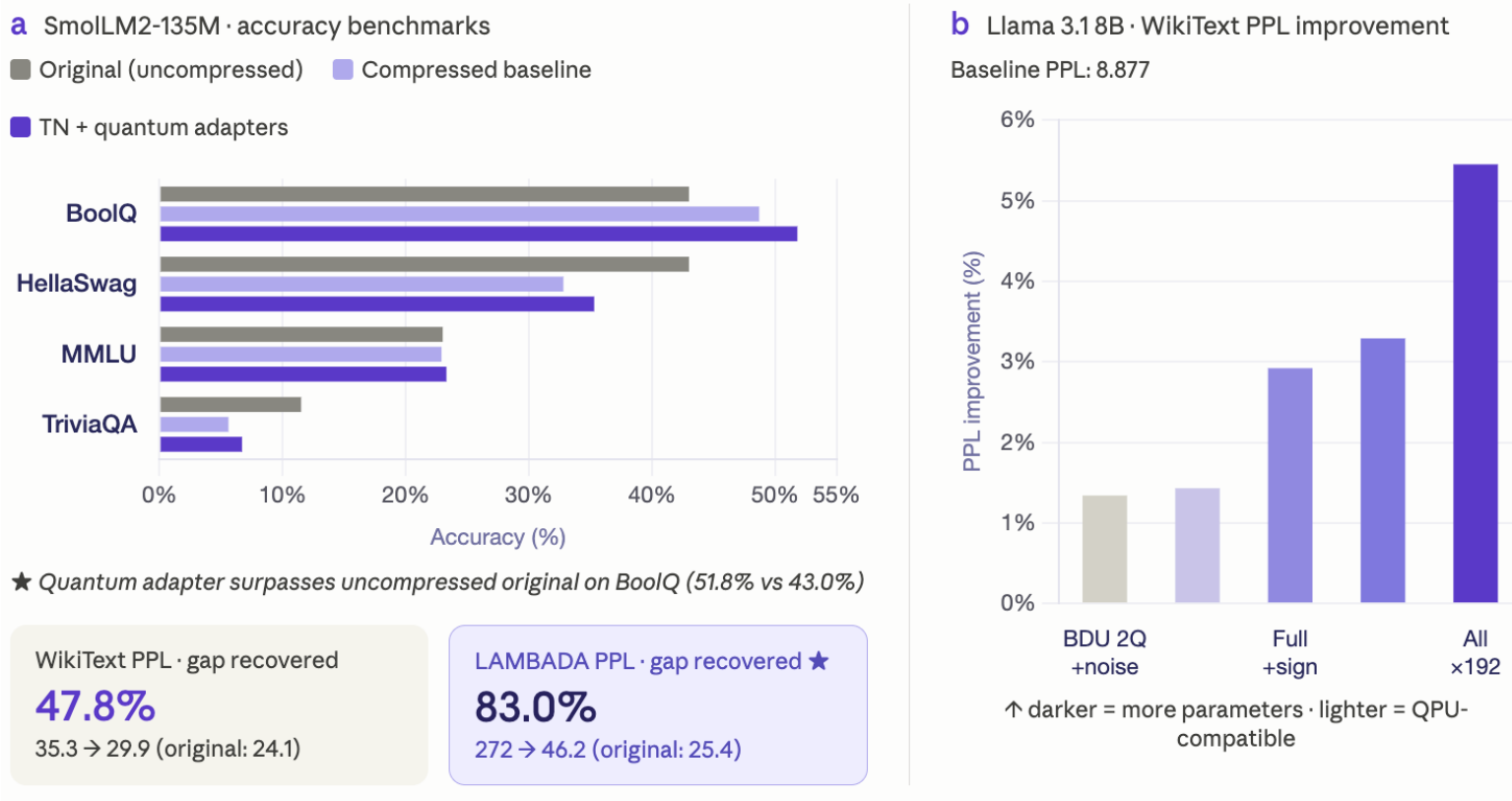}
  \caption{%
    \textbf{Multi-benchmark comparison across model scales.}
    \textbf{a,} Benchmark performance (WikiText PPL, LAMBADA PPL, BoolQ accuracy,
   HellaSwag normalised accuracy) for SmolLM2 (three configurations:
uncompressed original, compressed baseline, 210-layer CUA-enhanced).
\textbf{b,} Perplexity improvements for Llama~3.1~8B under different adapter
configurations (2-qubit BDU, full unitary + sign, unconstrained, all-sublayer BDU),
including noise simulation.
  }
  \label{fig5}
\end{figure}

\break

%% ================================================================
%% EXTENDED DATA
%% ================================================================

\setcounter{table}{0}

\section*{Extended Data}

\noindent\textbf{Extended Data Table 1 $|$ WikiText perplexity vs.\ block size
for 210-layer Cayley adapters.}
Three adapter regimes (unconstrained, orthogonal unitary, sign-constrained
unitary) for block dimensions from $4 \times 4$ ($n = 2$~qubits) to full
input dimension ($n = 9$--10~qubits), all applied to the compressed SmolLM2
backbone (94.8M parameters).
Sign-constrained unitaries are directly QPU-compatible.
Adapter parameter counts reflect unitary parameters only (backbone frozen).

\begin{table}
\centering
\caption*{}
\begin{tabular}{lccccccc}
\toprule
& & \multicolumn{2}{c}{Unconstrained} & \multicolumn{2}{c}{Unitary} & \multicolumn{2}{c}{Sign-constrained} \\
\cmidrule(lr){3-4}\cmidrule(lr){5-6}\cmidrule(lr){7-8}
$n$Q & Block & PPL & Params & PPL & Params & PPL & Params \\
\midrule
2 & 4     & 34.23 & 505K  & 34.73 & 189K  & 34.68 & 189K \\
3 & 8     & 33.79 & 1.01M & 34.36 & 442K  & 34.38 & 442K \\
4 & 16    & 33.25 & 2.02M & 33.91 & 948K  & 34.01 & 948K \\
5 & 32    & 32.47 & 4.04M & 33.34 & 1.96M & 33.50 & 1.96M \\
6 & 64    & 31.37 & 8.09M & 32.47 & 3.98M & 32.88 & 3.98M \\
7 & 128   & 30.15 & 17.5M & 31.47 & 8.66M & 32.14 & 8.66M \\
8 & 256   & 28.83 & 41.8M & 30.25 & 20.8M & 31.25 & 20.8M \\
9 & 384   & 27.57 & 104.3M & 28.83 & 52.0M  & 30.42 & 52.0M  \\
10 & 1024 & 27.01 & 220.2M & 27.97 & 110.0M & 30.05 & 110.0M \\
input & 384--896 & 27.07 & 79.7M & 27.97 & 39.8M  & 29.94 & 39.8M  \\
\midrule
\multicolumn{2}{l}{Compressed baseline} & \multicolumn{2}{c}{35.31} & \multicolumn{2}{c}{—} & \multicolumn{2}{c}{—} \\
\multicolumn{2}{l}{Original (teacher)} & \multicolumn{2}{c}{24.10} & \multicolumn{2}{c}{—} & \multicolumn{2}{c}{—} \\
\bottomrule
\end{tabular}
\end{table}

\bigskip
\noindent\textbf{Extended Data Table 2 $|$ Ablation study (layer 9
\texttt{up\_proj}).}
Comparison of learned Cayley adapter against stochastic baselines.
All configurations use a single-layer adapter on the compressed SmolLM2
backbone. Stochastic baselines report mean $\pm$ s.d.\ over 3 random seeds.

\begin{table}
\centering
\caption*{}
\begin{tabular}{lc}
\toprule
Adapter type & WikiText PPL \\
\midrule
Original SmolLM2-135M (no adapter) & 24.19 \\
Compressed baseline                & 35.21 \\
\midrule
Identity (no transform)            & 35.27 \\
Signed permutation (diag $\pm 1$)  & 35.27 $\pm$ 0.00 \\
Random Gaussian (non-orthogonal)   & 38.38 $\pm$ 0.26 \\
Random unitary                     & 38.93 $\pm$ 0.45 \\
Random permutation                 & 41.59 $\pm$ 0.53 \\
\midrule
\textbf{Learned Cayley (ours)}     & \textbf{34.89} \\
\bottomrule
\end{tabular}
\end{table}

\bigskip
\noindent\textbf{Extended Data Table 3 $|$ Multi-benchmark evaluation for
SmolLM2.}
Zero-shot performance of three configurations on five standard benchmarks.
LAMBADA reports perplexity (lower is better); all others report accuracy
(higher is better). ``TN 210L'': compressed backbone + 210-layer
sign-constrained Cayley unitary adapters.

\begin{table}
\centering
\caption*{}
\begin{tabular}{lccc}
\toprule
Benchmark & Original & Compressed & TN 210L \\
\midrule
WikiText PPL $\downarrow$  & 24.10 & 35.29 & \textbf{29.94} \\
LAMBADA PPL $\downarrow$   & 25.42 & 272.18 & \textbf{46.20} \\
BoolQ Acc $\uparrow$       & 43.0\% & 48.7\% & \textbf{51.8\%} \\
HellaSwag Acc$_{\rm norm}$ $\uparrow$ & 43.0\% & 32.8\% & 35.3\% \\
MMLU Acc $\uparrow$        & 23.0\% & 22.9\% & 23.3\% \\
TriviaQA Acc $\uparrow$    & 11.5\% & 5.6\%  & 6.7\% \\
\bottomrule
\end{tabular}
\end{table}

\bigskip
\noindent\textbf{Extended Data Table 4 $|$ Noise scalability across qubit counts
under IBM Heron r2 error rates.}
Per-gate depolarising simulation + readout + multinomial shot noise
($N_{\rm shots} = 1{,}024$), for 210-layer sign-constrained Cayley adapters.
$\lambda_{1Q}$ and $\lambda_{2Q}$: effective depolarising parameters from
single-qubit (SX) and two-qubit (CZ) gates; $\lambda$: total gate infidelity;
$\varepsilon_{\rm ro}$: $n$-qubit readout error.

\begin{table}
\centering
\caption*{}
\small
\begin{tabular}{lcccccccc}
\toprule
$n$Q & Block & SX & CZ & $\lambda_{1Q}$ & $\lambda_{2Q}$ & $\lambda$ & $\varepsilon_{\rm ro}$ & PPL (noisy) \\
\midrule
2 & 4   & 20    & 4     & 0.005 & 0.007 & 0.012 & 0.014 & 45.50 \\
3 & 8   & 188   & 45    & 0.045 & 0.077 & 0.119 & 0.020 & 1,189 \\
4 & 16  & 992   & 273   & 0.216 & 0.385 & 0.518 & 0.027 & $2.3 \times 10^6$ \\
5 & 32  & 4,526 & 1,331 & 0.671 & 0.907 & 0.969 & 0.034 & $4.9 \times 10^7$ \\
6 & 64  & 18,708 & 5,534 & 0.990 & $\approx$1 & $\approx$1 & 0.040 & $4.8 \times 10^7$ \\
7 & 128 & 75,804 & 22,396 & 1.0 & 1.0 & 1.0 & 0.047 & $5.7 \times 10^7$ \\
8 & 256 & 306,846 & 91,000 & 1.0 & 1.0 & 1.0 & 0.053 & $8.2 \times 10^7$ \\
\midrule
\multicolumn{3}{l}{Noiseless (2Q)} & & & & & & 34.68 \\
\bottomrule
\end{tabular}
\end{table}

\bigskip
\noindent\textbf{Extended Data Table 5 $|$ Llama~3.1~8B perplexity
under different adapter configurations.}
All results use the \texttt{ibm\_basquecountry} backend (2-qubit BDU circuits
unless otherwise stated). ``Improvement'' is relative to the unmodified base model
(PPL 8.877). BDU+noise: classical simulation with IBM Heron r2 per-gate
depolarising channels.

\begin{table}
\centering
\caption*{}
\begin{tabular}{lccc}
\toprule
Configuration & PPL & Extra params & Improvement \\
\midrule
Llama-3.1-8B-Instruct (baseline)    & 8.877 & --- & --- \\
+ 2-qubit BDU (1 layer, v\_proj 7)  & 8.752 & 6K & +1.43\% \\
+ 2-qubit BDU + noise               & 8.759 & 6K & +1.34\% \\
+ Full unitary + sign (1 layer)     & 8.618 & 8.38M & +2.92\% \\
+ Unconstrained (1 layer)           & 8.585 & 16.77M & +3.29\% \\
+ 2-qubit BDU (all sublayers, 192 adapters) & 8.393 & --- & +5.45\% \\
\bottomrule
\end{tabular}
\end{table}

\bigskip
\noindent\textbf{Extended Data Table 6 $|$ Representative qualitative examples
from real QPU inference (ibm\_basquecountry, IBM Heron r2).}
The QPU executes all 144 two-qubit unitary blocks for every token.
Responses are verbatim from QPU inference.
$\dagger$: original SmolLM2-135M also answers correctly.

\begin{table}
\centering
\caption*{}
\begin{tabular}{llll}
\toprule
Prompt & Original & Compressed & QPU output \\
\midrule
$9 \times 6$?  & ``54 $\times$ 6 = 312'' & ``6 $\times$ 6 = 36'' & ``9 times 6 is 54'' \checkmark \\
$2 + 2$?$^\dagger$ & \checkmark & Rambles, never says 4 & ``we get 4'' \checkmark \\
$2 + 9$?$^\dagger$ & \checkmark & ``2 + 3 = 5''          & ``2 + 9 = 11'' \checkmark \\
Largest organ?$^\dagger$ & \checkmark & ``The heart''     & ``The liver'' \checkmark \\
\bottomrule
\end{tabular}
\end{table}

\bigskip
\noindent\textbf{Extended Data Table 7 $|$ Entanglement characterisation of
trained full-dimension Cayley unitaries.}
For each unitary dimension, worst-case (highest) average effective bond dimension
across all bipartitions, maximum effective bond dimension, and average entropy
ratio (fraction of maximum possible entanglement entropy for a Haar-random
unitary). 210 unitaries total.

\begin{table}
\centering
\caption*{}
\begin{tabular}{lccccc}
\toprule
Dim. & $n$Q & Count & Avg.\ eff.\ bond dim. & Max eff.\ bond dim. & Entropy ratio \\
\midrule
384  & 9  & 22  & 84  & 153 & 4.7\% \\
576  & 10 & 158 & 114 & 255 & 5.6\% \\
896  & 10 & 30  & 201 & 256 & 7.8\% \\
\bottomrule
\end{tabular}
\end{table}

\bigskip
\noindent\textbf{Extended Data Table 8 $|$ Qualitative examples where the
210-layer CUA-enhanced SmolLM2 surpasses both classical baselines.}
All examples are ``TN beats both'' wins: the CUA-enhanced model answers correctly
while both the uncompressed original and the compressed baseline fail.

\begin{table}
\centering
\caption*{}
\begin{tabular}{llll}
\toprule
Question & Original & Compressed & TN Model (ours) \\
\midrule
What is $2+2$?  & Rambles, never says 4 & Rambles, never says 4 & ``result 4'' \checkmark \\
$7 \times 6$?   & ``6 $\times$ 7 = 36'' & ``6 $\times$ 6 = 36'' & ``42'' \checkmark \\
$7 \times 4$?   & ``7 $\times$ 16 = 96'' & ``8 $\times$ 4 = 32'' & ``28'' \checkmark \\
$13 + 6$?       & ``13 + 6 = 29'' & ``Sum = 130'' & ``19'' \checkmark \\
$22 - 14$?      & ``22 $- 14 = 10$'' & ``22 = 12'' & ``22 $-$ 14 = 8'' \checkmark \\
$\sqrt{169}$?   & Never computes & ``$\approx$10.42'' & ``13'' \checkmark \\
\bottomrule
\end{tabular}
\end{table}

\bigskip
\noindent\textbf{Extended Data Table 9 $|$ Operator Schmidt decomposition of the Llama-3.1-8B \texttt{v\_proj}-7 BDU adapter.}
The trained $4{,}096{\times}4{,}096$ block-diagonal unitary ($1{,}024$ SO(4) blocks) is reinterpreted as a $12$-qubit unitary. Operator Schmidt decomposition is computed across the natural block-index$|$intra-block bipartition (10$|$2 cut) and across the equal-mass bipartition (6$|$6 cut). ``Rank max'' is the maximum operator Schmidt rank permitted by the cut. Rescaling the Cayley angles by $s$ stress-tests the structural capacity of the family.
\begin{table}[h!]
\centering
\caption*{}
\small
\begin{tabular}{llcccc}
\toprule
Object & Bipartition & Rank max & OSR achieved & $\sigma_{\max}$ & $1{-}\sum\sigma_i^4$ \\
\midrule
CUA (trained)              & natural 10$|$2 & 16   & 16 (saturated) & 0.953 & 0.174 \\
CUA (trained)              & equal 6$|$6    & 4096 & 64             & 0.953 & 0.174 \\
Identity                   & --             & --   & 1              & 1.000 & 0.000 \\
Haar SO(4) BDU             & natural 10$|$2 & 16   & 16             & --    & 0.937 \\
Depth-1 VQC brickwork      & natural 10$|$2 & 16   & 1              & --    & 0.000 \\
\midrule
Stress, $s{=}0$            & natural 10$|$2 & 16   & 1              & 1.000 & 0.000 \\
Stress, $s{=}1$ (trained)  & natural 10$|$2 & 16   & 16             & 0.953 & 0.174 \\
Stress, $s{=}4$            & natural 10$|$2 & 16   & 16             & 0.547 & 0.855 \\
Stress, $s{=}10$           & natural 10$|$2 & 16   & 16             & 0.323 & 0.924 \\
\bottomrule
\end{tabular}
\end{table}

\break 

%% ================================================================
%% END MATTER
%% ================================================================

\section*{Acknowledgements}

We acknowledge the Donostia International Physics Center (DIPC), Ikerbasque
Foundation for Science, Basque Government, Diputación de Gipuzkoa, European
Innovation Council (EIC), Tecnun and Spanish Government for constant support.
We thank the IBM Quantum Network for access to the IBM System Two processors
(\texttt{ibm\_basquecountry} and \texttt{ibm\_strasbourg}).

\section*{Author contributions}

Conceptualisation: R.O., B.A.
Methodology: B.A., S.S., S.S.J., A.K.
Software and quantum circuit implementation: B.A.
Formal analysis and data curation: B.A., S.S.
Writing (original draft): B.A., R.O.
Writing (review and editing): all authors.
Supervision: R.O.
Project administration and funding acquisition: R.O.

\section*{Competing interests}

The authors declare no competing interests.

\section*{Additional information}

\noindent\textbf{Extended Data} is available for this paper under request.\\
\noindent\textbf{Correspondence} and requests for materials should be
addressed to Román Orús \\
(\texttt{roman.orus@multiversecomputing.com}).

\end{document}